\documentclass[aps,preprint,preprintnumbers,amsmath,amssymb,floatfix,showpacs]{revtex4}
\usepackage[cp1251]{inputenc}
\usepackage{graphicx}
\usepackage{epsfig}

\usepackage{amsbsy,bm}

\begin{document}

\title{Interpretation of the time delay in the ionization of Coulomb and two-center systems}

\author{Vladislav V. Serov, Vladimir L. Derbov, Tatyana A. Sergeeva}
\affiliation{
Department of Theoretical Physics, Saratov State University, 83
Astrakhanskaya, Saratov 410012, Russia}

\date{\today}
\begin{abstract}
We consider the time delay of electron detachment from a Coulomb center and two-center systems in the process of ionization. It is shown that the attosecond streaking, most usual method of time delay measure, can be formally described by placing a virtual detector of the arrival time delay at a certain distance from the center of the system. This approach allows derivation of a simple formula for Coulomb-laser coupling that perfectly agrees with the results of numerical solution of the time-dependent Schr\"odinger equation. The dependence of the time delay upon the energy, the angular momentum projection, and the azimuthal quantum number is studied for the ionization of molecular hydrogen ion. Finally, we propose a physical interpretation of singularities, arising when the formal expression for the time delay is applied to the ionization of molecular hydrogen.
\end{abstract}

\pacs{33.80.Eh, 32.80.Fb, 33.20.Xx}
\maketitle

\section{Introduction}
Recently the appearance of laser systems that can generate super-intense pulses as short as a few hundreds of attoseconds gave rise to new capabilities in studying electron dynamics in atoms and molecules, which were not available earlier. Among these of particular interest are the methods of measuring the delay of electron ejection from the atom subject to photoionization   \cite{ScrinziMIvanov2006}. At present the measurements of ionization delay are already performed in noble gas atoms using the methods of attosecond streaking \cite{Scrinzi2010} and RABITT (reconstruction of attosecond beating by interference of two-photon transition) \cite{Klunder2011}. 
The time delays in photoionization of molecules are a subject of growing theoretical interest (see \cite{IvanovSerov2012} and references therein). %

\begin{figure}[ht]
\begin{center}
\includegraphics[angle=-90,width=0.5\textwidth]{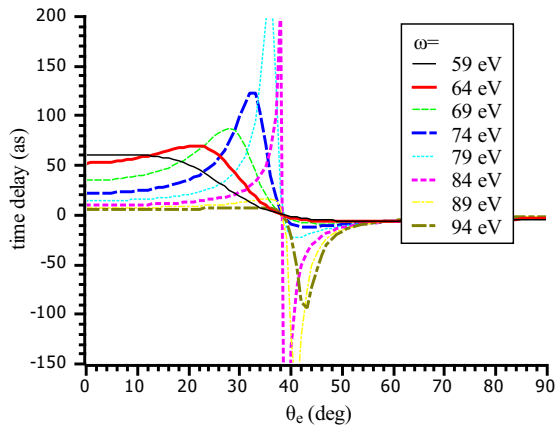}
\\(a)\\
\includegraphics[angle=-90,width=0.5\textwidth]{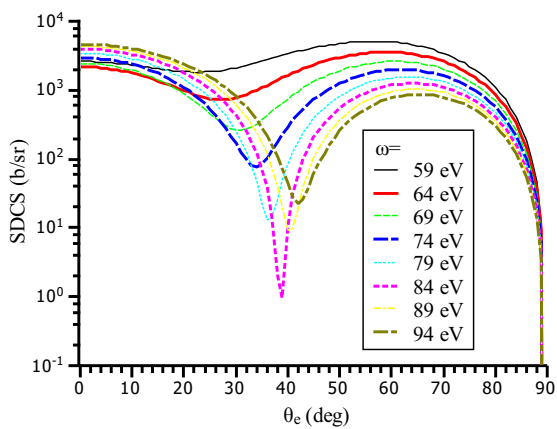}\\(b)
\end{center}
\caption{(Color online) (a) Wigner delay and (b) single differential cross-section  $\frac{d\sigma}{d\Omega_e}$ of the single ionization of molecular hydrogen  H$_2$ versus the ejection angle  $\theta_e$ at $\mathbf{R}\parallel\mathbf{e}$ for different values of the ejected electron energy.\label{FIG_H2_Wigner_delay}}
\end{figure}

The numerical study \cite{IvanovSerov2012} of the delayed electron ejection in the process of single-photon ionization of a hydrogen molecule H$_2$ revealed an interesting feature, namely, at certain energies of the ejected electron the angular distribution of the molecular photoionization time delay possesses singularities. Fig. \ref{FIG_H2_Wigner_delay}a shows the Wigner time delay $t_W$ vs the electron ejection angle $\theta_e$ for single ionization of oriented  H$_2$ molecule under the condition that the molecular axis is parallel to the polarization of radiation, $\mathbf{R}\parallel\mathbf{e}$. The dependence was calculated for different energies of the ejected electron using the external complex scaling (ECS) method \cite{Serov2009}. For the energy of ionizing radiation photon  $\hbar\omega=84$ eV there is a discontinuity at the angle $\theta_e=39^\circ$, in the vicinity of which  $t_W$ tend to $\pm\infty$. From Fig.\ref{FIG_H2_Wigner_delay}b one can see that this discontinuity coincides with the zeros of the ionization differential cross-section. Although the probability of electron ejection in the directions close to this zero is small, it is still  possible in principle to measure the time delay for the electron ejection in these directions. Large negative values of  $t_W$ contradict the causality principle  \cite{Wigner1955}, since formally they mean that the electron is ejected by the molecule long before the absorption of the photon.  

The aim of the present paper is to clarify the conditions that cause singularities in  $t_W$ and to resolve the paradox of large negative delay in two-center systems. We consistently reconsidered the physical meaning of Wigner time delay $t_W$ for Coulomb systems, studied the dependences of $t_W$ upon the energy and the ejection angle by the example of the ionization of the model two-center system (the molecular ion  H$_2^+$), and analyzed the dynamics of the ejected electron wave packet under the conditions, for which the singularities of $t_W$ take place.

The paper is organized as follows. In Section  \ref{sec:SingleCoulomb} the general question of choosing the origin for the time delay in the case of Coulomb field. In Section  \ref{sec:streakingCoulomb} we demonstrate that the known method of attosecond streaking is  equivalent to a device, located at a certain distance from the center and detecting the delay of the electron arrival, and that the Coulomb-laser coupling  \cite{Smirnova2007,Smirnova2011}, arising in the theory of the attosecond streaking, is due to the Coulomb  advance of the electron arrival at this virtual device. In Section \ref{sec:TwoCoulomb} we study the energy dependence of the time delay for spheroidal Coulomb waves.  Section  \ref{sec:H2plus} is devoted to the origin of singularities in the angular distribution of the Wigner time delay in two-center targets, and Section \ref{sec:Gaussians} presents the physical interpretation of these singularities.  

Below we use the atomic units of measurement, except other declared.

\section{Time delay in one-center system with Coulomb potential}\label{sec:SingleCoulomb}

In the framework of quantum mechanics the behavior of a system, ionized by an external laser field, is described by the wave function. After the external action is finished, it can be presented in the form    
\begin{eqnarray}
\psi(\mathbf{r},t)=\int f(\mathbf{k})\varphi^{(-)}_{\mathbf{k}}(\mathbf{r})e^{-i\frac{k^2}{2}t}d\mathbf{k},
\end{eqnarray}
where $f(\mathbf{k})$ is the ionization probability amplitude, $\varphi^{(-)}_{\mathbf{k}}(\mathbf{r})$ is the wave function of the continuous spectrum, describing the free particle with the momentum $\mathbf{k}$ at the infinitely large distance.

If the potential of interaction of the particle with the center is short-range, then $\varphi^{(-)}_{\mathbf{k}}(\mathbf{r}\to\infty)=(2\pi)^{-3/2}\exp(i\mathbf{k}\mathbf{r})$.
Using the known relation
\[\lim_{r\to\infty}e^{ikr\mathbf{n}\mathbf{n}'}=\frac{2\pi}{ikr}\left[e^{ikr}\delta(\mathbf{n}-\mathbf{n}')-e^{-ikr}\delta(\mathbf{n}+\mathbf{n}')\right] \]
(where $\delta(\mathbf{n}+\mathbf{n}')$ is a delta-function), 
we get that for  $r\to\infty$ the wave function can be expressed as  
\begin{eqnarray}
\psi(\mathbf{r},t)\sim \int_0^\infty  |f(k\mathbf{n})|e^{iS_+}dk
-\int_0^\infty |f(-k\mathbf{n})|e^{iS_-}dk. \label{asympt_psi}
\end{eqnarray}
Here the phases of the integrands 
\begin{eqnarray*}
S_{\pm}=\pm kr+\delta(\pm k\mathbf{n})-\frac{k^2}{2}t,
\end{eqnarray*}
where $\delta(\mathbf{k})=\arg f(\mathbf{k})$ is the phase of the ionization complex amplitude. In the limit $t\to\infty$ the major contribution to the integrals is introduced by the vicinity of the stationary points  $k_0=k_0(\mathbf{r},t)$, in which the derivative of the phase of the integrand is equal to zero,
\begin{eqnarray*}
\left.\frac{dS_{\pm}}{dk}\right|_{k_0}=\pm r+\left.\frac{\partial\delta(\pm k\mathbf{n})}{\partial k}\right|_{k_0}-k_0t=0,
\end{eqnarray*}
since far from the points  $k_0$ the integrands are fast-oscillating.  For $k\geq 0$ only $S_{+}$ possesses an extremum, so that only the first term in Eq. (\ref{asympt_psi}) will contribute to the wave function, and 
\(
\psi(\mathbf{r},t)\sim f(k_0\mathbf{n})e^{ik_0r-i\frac{k_0^2}{2}t}.
\)
This fact can be interpreted as follows. At the point  $\mathbf{r}=r\mathbf{n}$ at the moment of time
\begin{eqnarray}
t=\frac{r}{k}+\frac{1}{k}\frac{\partial\delta(\mathbf{k})}{\partial k}
\end{eqnarray}
one can detect the particle having the momentum $\mathbf{k}=k\mathbf{n}$ with the maximal probability. Since the ratio  $r/k$ is the time, required for the arrival at the point  $\mathbf{r}$ of the particle that left the center $r=0$ at the time $t=0$ and moved with the uniform velocity  $k$, the expression
\begin{eqnarray}
t_W=\frac{1}{k}\frac{\partial\delta(\mathbf{k})}{\partial k}=\frac{\partial\delta}{\partial E}(\mathbf{k}).
\end{eqnarray}
has the physical meaning of the time delay of the particle arrival at the distance $r$ from the center with respect to that in the case of uniform rectilinear motion.  Note that the condition that the particle should be ejected directly from the center is not necessary, because if the particle has the impact parameter $b$ (which in the case of free motion is the minimal distance between the particle and the origin), then $\lim_{t\to\infty}(r/k)=\lim_{t\to\infty}\sqrt{b^2+(kt)^2}/k=t$.

The interpretation of the energy derivative of the phase as the  time delay was first proposed by Eisenbud and Wigner. In \cite{Wigner1955} Winger has shown that if a particle falls from infinity, then after passing the center of a short-range potential the delay with respect to the motion in free space equals to $2t_W$. In the same paper he has shown that if the potential is attractive, then $t_W>0$ only for near-resonance energy (i.e., when the particle is captured into a quasistationary state for a long time). For the rest values of energy $(-r_{pot}/k)\leq t_W \leq 0$ (where $r_{pot}$ is the potential action radius), i.e., corresponds to the expected from the point of view of classical physics. For brevity below we will refer the energy derivative of the phase, $t_W$, as Wigner time delay.

One more important note should be made before proceeding further. If the external impact on the system is weak, so that one can calculate the ionization amplitude using the first-order perturbation theory, then  
\begin{eqnarray}
f(\mathbf{k})=\langle \mathbf{k}|\hat{w}|i\rangle,\label{ampl1B}
\end{eqnarray}
where $|i\rangle$ is the wave function of the initial state of the system, $|\mathbf{k}\rangle\equiv\varphi^{(-)}_{\mathbf{k}}(\mathbf{r})$ is the wave function of the final state, describing the free electron with the momentum  $\mathbf{k}$ at the infinity, $\hat{w}$ is the operator of external perturbation. 

As known, for a centrally symmetrical system the wave function is representable is the form of expansion
\begin{eqnarray}
\varphi^{(-)}_{\mathbf{k}}(\mathbf{r})=4\pi\sum_{\ell m}i^\ell e^{-i\delta_{\ell}(k)}Y_{\ell m}^*(\widehat{k})Y_{\ell m}(\widehat{r})R_{k\ell}(r),
\end{eqnarray}
where 
$R_{k\ell}(r\to\infty)=\frac{1}{kr}\sin(kr-\pi\ell/2+\delta_\ell)$ is the radial wave function of the partial spherical wave, $\delta_\ell$ is the phase of the spherical wave. In this case the amplitude is expressed as
\begin{eqnarray}
f(\mathbf{k})=\sum_{\ell m}A_{\ell m}(k)i^{-\ell} e^{i\delta_{\ell}(k)}Y_{\ell m}(\widehat{k}).
\end{eqnarray}
If the transition amplitude  $A_{\ell m}=4\pi\langle R_{k\ell} Y_{\ell m} | \hat{w} |i\rangle$ differs from zero for only one  $\ell$, then, obviously, the Wigner time delay
 depends only on the corresponding partial phase
\begin{eqnarray}
t_W=\frac{d\delta_{\ell}}{dE}. \label{tauW_ell}
\end{eqnarray}
In other words, if in an one-center system the transition occurs into a state with fixed angular momentum, then the Wigner time delay appears to be completely determined by the energy and the quantum number of the squared angular momentum, and is by no means related
to the initial state and the particular form of the perturbation potential.   

If the potential is non short-range and tends to a Coulomb one, $Z/r$, at large distances, then the asymptotic form of the wave function becomes essentially different, since the phase in Eq.(\ref{asympt_psi}) will contain the term logarithmic in $r$, taking the form 
\begin{eqnarray}
S_+=kr+\frac{Z}{k}\ln 2kr+\delta(k\mathbf{n})-\frac{k^2}{2}t.
\end{eqnarray}
The stationary point will be now determined by the solution of the equation 
\begin{eqnarray}
\frac{dS_+}{dk}=r-\frac{Z}{k^2}\ln 2kr+\frac{Z}{k^2}+\frac{d\delta}{dk}-kt=0.
\end{eqnarray}
From this it is seen that the time of arrival of the electron, having the momentum  $k$, at the point at the distance  $r$ from the Coulomb center with the charge  $Z$ will be equal to
\begin{eqnarray}
t(r)=\frac{r}{k}-\frac{Z}{k^3}\ln 2kr+\frac{Z}{k^3}+t_{W}.\label{t_r_QuantumCoulomb}
\end{eqnarray}

As it has been already mentioned, in systems with short-range potentials the Wigner time delay has simple physical meaning, namely, it is the delay of arrival of the particle, ejected from the force center, as compared to the case of rectilinear and uniform motion. To understand the physical meaning of Wigner time delay in the case of Coulomb field, the comparison should be made with a certain classical motion in Coulomb field. The most rational choice is the motion of a particle with zero angular momentum, starting from the center  $r=0$ at the moment of time $t=0$. The law of such motion can be obtained from the known general law of a particle motion with arbitrary angular momentum.  
However, it is simpler to use the known general formula for the law of the motion in one-dimensional case, 
\begin{eqnarray}
t_{C}(r)&=&\int_0^{r}\frac{dr}{p(r)}, \label{tr_common}
\end{eqnarray}
where $p(r)=\sqrt{k^2+2Z/r}$ is the momentum of the particle at the distance $r$ from the center. Performing the integration, we get
\begin{eqnarray}
t_{C}(r)=\frac{p(r)r}{k^2}+\frac{Z}{k^3}\ln\frac{p(r)/k-1}{p(r)/k+1}.
\end{eqnarray}
The asymptotic form of this law is
\begin{eqnarray}
t_{C}(r\to\infty)=\frac{r}{k}-\frac{Z}{k^3}\ln\frac{2k^2r}{Z}+\frac{Z}{k^3}.\label{tC_for_r_infty)}
\end{eqnarray}
Let us introduce the notion of the delay of a "`quantum"' particle moving in the asymptotic Coulomb field with respect to a classical particle with the angular momentum  $\ell=0$, ejected from the center at the time moment $t=0$
\begin{eqnarray}
t_0=\lim_{r\to\infty}[t(r)-t_{C}(r)]. \label{eq_t0}
\end{eqnarray}
From Eqs. \eqref{t_r_QuantumCoulomb} and \eqref{tC_for_r_infty)} it is clear that the ejection delay related to Wigner time delay as %
\begin{eqnarray}
t_0=t_{W}+\frac{Z}{k^3}\ln\frac{k}{Z}. \label{eq_t0ln} 
\end{eqnarray}

\begin{figure}[ht]
\begin{center}
\includegraphics[angle=-90,width=0.5\textwidth]{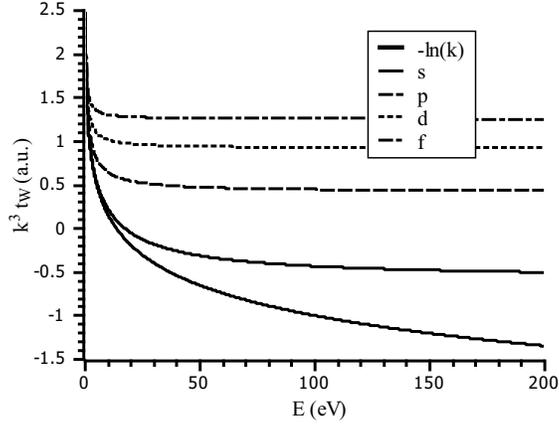}\\(a)\\
\includegraphics[angle=-90,width=0.5\textwidth]{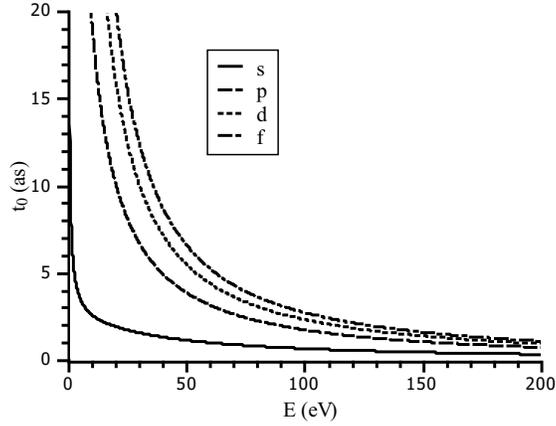}\\(b)
\end{center}
\caption{Dependence of the time delay upon the energy of the ejected electron for a hydrogen atom: a)Wigner time delay; b) the ejection delay, Eq. (\ref{eq_t0ln}) \label{FIG_H_delay}.}
\end{figure}

Fig. \ref{FIG_H_delay}a shows the Wigner time delays for the continuum of a hydrogen atom at different values of the angular momentum $\ell$, mutiplied by  $k^3$ for clearness. It is seen that at small energies $t_W\to -\frac{Z}{k^3}\ln\frac{k}{Z}$ for all values of  $\ell$. In Fig.\ref{FIG_H_delay}b the delays  $t_0$ are shown. At $k\to 0$ they tend to infinity, but not as $-\frac{Z}{k^3}\ln\frac{k}{Z}$, but much slower, as  $1/k^2$. Besides, $t_0>0$ for all values $E$ and grows with increasing $\ell$. This can be interpreted as follows. The centrifugal potential can be considered as a short-range repulsive one, and the particle motion through it results in the delay with respect to the motion in the pure Coulomb field. 

Let us also consider the time delay %
for the particle, arriving at the detector located at the distance $r$ from the center, in comparison with the case of uniform and rectilinear motion  
\begin{eqnarray}
t_{D}(r)=t_0+t_C(r)-\frac{r}{k}\simeq t_0-\frac{Z}{k^3}\ln\frac{2k^2r}{Z}+\frac{Z}{k^3}.\label{eq_tD}
\end{eqnarray}
Obviously, at large  $r$, when the logarithmic term, describing the Coulomb advance
\begin{equation}
t_{CA}(r)=t_C(r)-\frac{r}{k}\simeq -\frac{Z}{k^3}\left[\ln\frac{2k^2r}{Z}-1\right],
\end{equation}
becomes dominant, the time delay will be negative.

\section{Physical meaning of Coulomb-laser coupling in attosecond streaking}\label{sec:streakingCoulomb}

Apparently, the above speculations %
of a detector, providing direct measurement of the particle arrival delay, concerns a gedanken experiment rather that a real one.  In real experiments the delays are recorded using indirect approaches,  the attosecond streaking  \cite{Scrinzi2010} and the reconstruction of attosecond beating by interference of two-photon transition (RABITT) \cite{Klunder2011}. Is is interesting to relate the above discussion with these real measurements.  In the present Section we demonstrate that the correction to the attosecond streaking time delay measurement, referred to as  Coulomb–laser coupling \cite{Smirnova2007}, in fact is nothing but the Coulomb advance $t_{CA}(r)$, accumulated to the moment of essential variation of the laser field strength. This actually means that the attosecond streaking is exactly equivalent to a virtual detector, placed at a fixed distance from the center and measuring the time delay (\ref{eq_tD}) of the particle arrival.

In the method of attosecond streaking \cite{Itatani2002}
the atom is simultaneously affected by the extreme ultraviolet (XUV) ionizing pulse with the duration of about 100 attoseconds and the auxiliary variable field in the form of a femtosecond IR laser pulse. The latter contributes to the momentum of the ejected electron, depending upon the time of ejection, which makes it possible to measure the ejection delay.     

Let the electron be a subject to the auxiliary field having the strength $\mathbf{\mathcal{E}}(t)$. The potential energy of the electron in such field is  
\begin{eqnarray}
V(\mathbf{r},t)=\mathbf{\mathcal{E}}(t)\cdot\mathbf{r}.
\end{eqnarray}
The detection of electrons is usually implemented
in the direction of the polarization vector of the auxiliary field, since in such geometry of the experiment the auxiliary field does deflect the electron trajectory. Since during the action of the IR field the electron has a time to get far enough from the center compared to the atomic radius, the angular motion can be neglected. Thus, we get one-dimensional problem of an electron moving in the potential   
\begin{eqnarray}
V(r,t)=\mathcal{E}(t)r.
\end{eqnarray}

We start from the case hard to realize experimentally, but more transparent for theoretical consideration, the constant IR field, suddenly switched off at the time  $T$:
\begin{eqnarray}
\mathcal{E}(t)=\left\{ \begin{array}{ll}
\mathcal{E}_0, & t<T;\\
0, & t>T.
\end{array} \right. \label{step_Et}
\end{eqnarray}
We also assume the strength of this field  $\mathcal{E}_0$ to be very small. Note, that although the practical detection of the momentum increment requires rather strong IR fields, no fundamental limitations follow from this requirement. The attosecond streaking works by virtue of the first-order effect surviving even for an infinitesimally weak field, so the assumption of the smallness of the $\mathcal{E}_0$ magnitude is quite valid. %

At the time $\tau$ the atom is affected by an ultrashort pulse of XUV radiation. As a result an electron is ejected with the energy  $E$, equal to the difference of the mean energy of the XUV photon $\omega_0$ and the ionization potential  $I$. Until the auxiliary IR field is unchanged, the energy will obviously be conserved, so that we can write the  conservation law as  
$E=p^2(t)/2+\mathcal{E}_0r(t)$. From the latter we can express the value of the additional momentum, acquired before the field is switched off: $\Delta p=-\mathcal{E}_0r(T)/p(T)$. If in this formula we set  $r(T)\simeq r^{(0)}(T-\tau)$ and $p(T)\simeq p^{(0)}(T-\tau)$, i.e., instead of the exact position and momentum we use their values for the particle that appeared in the center at the time  $\tau$ and moves in the absence of the auxiliary IR field, then only the second-order terms in $\mathcal{E}_0$ will be changed. In the first-order approximation with respect to  $\mathcal{E}_0$ we obtain
\begin{eqnarray}
\Delta p=-\mathcal{E}_0\frac{r^{(0)}(T-\tau)}{p^{(0)}(T-\tau)}.
\end{eqnarray}
Both in systems with short-range potential and in Coulomb systems it is possible to assume $p^{(0)}(T-\tau)\approx k$ at sufficiently large $T-\tau$. For systems with short-range potential  $r^{(0)}(T-\tau)\approx k(T-\tau-t_W)$ and, correspondingly,
\begin{equation}
\Delta p=-\mathcal{E}_0(T-\tau-t_W).
\end{equation}
Thus, in the plot of the dependence $\Delta p(\tau)$ the time of ejection will manifest itself as a horizontal shift of the plot with respect to the straight line $-\mathcal{E}_0(T-\tau)$.

In systems with Coulomb potential $r^{(0)}(T-\tau)\approx k\{T-\tau-t_{D}[r^{(0)}(T-\tau)]\}$. Since $t_{D}$ (see Eq.(\ref{eq_tD})) logarithmically depends upon the distance, one can use  $t_{D}[r^{(0)}(T-\tau)]\approx t_{D}[k(T-\tau)]\}$, which yields
\begin{equation}
\Delta p=-\mathcal{E}_0\{T-\tau-t_{D}[k(T-\tau)]\}.\label{Dp_C_step_Et}
\end{equation}
Thus, the attosecond streaking works as a detector, placed at the distance  $r_{eff}=k(T-\tau)$ from the point of free particle generation. 

We would like to emphasize the following feature of the approach leading to this conclusion. Although the particle changes its velocity in the time of the field action, starting the acceleration directly after the transition to the unbound state, the energy conservation law allows expression of this change in terms of the particle behavior far from the center, so that the unknown details of the particle behavior near the center become unimportant. It may be said that until the field is switched on, the variation of the momentum is a dynamical variable that takes the fixed value only at the time moment when the field is switched off.  

Now consider the field strength $\mathcal{E}(t)$ smoothly changing with time. The variation of the particle total energy in time obeys the equation  $\frac{dE}{dt}(t)=\frac{\partial U}{\partial t}(r(t),t)$. From here it is easy to get that the first-order correction with respect to  $\mathcal{E}(t)$ to the particle energy obeys the equation 
 $\frac{d(\Delta E)}{dt}=\frac{d\mathcal{E}}{dt}(t)r^{(0)}(t-\tau)$. On the other hand, if the natural condition  $\mathcal{E}(t\to\infty)=0$ holds, then $\Delta E(t\to\infty)=k\Delta p$.
From here we get the increment of the momentum   
\begin{eqnarray}
\Delta p
=\int_\tau^\infty\frac{d\mathcal{E}}{dt}\frac{r^{(0)}(t-\tau)}{k}dt.
\end{eqnarray}
The distance $r^{(0)}(t-\tau)$ between the center and the particle is small at small $t-\tau$. With time the distance $r^{(0)}$ unlimitedly grows, therefore, at $t=\tau$ the integrand may be expected to contribute to the integral insignificantly. However, the subject of our interest is just the contribution to $r^{(0)}(t-\tau)$ from the shift that does not grow with time. Hence, for sure we assume that  $\frac{d\mathcal{E}}{dt}(t\to\tau)\to 0$. Below we will demonstrate that this assumption does not impede the applicability of this approach for realistic dependences $\mathcal{E}(t)$). Since in this case $r^{(0)}(t-\tau)$ does not contribute to the integral at small $t-\tau$, for short-range potentials we again can use the approximation  $r^{(0)}(t-\tau)\simeq k(t-\tau-t_W)$. Integrating by parts and using the inverse Taylor expansion, we get  
\begin{eqnarray}
\Delta p&=&
\int_\tau^{\infty}\frac{d\mathcal{E}}{dt}(t)\{t-\tau-t_W\}dt=\mathcal{E}(\tau)t_W-\int_\tau^{\infty}\mathcal{E}(t)dt\nonumber\\
&=&-\mathcal{A}(\tau)-\frac{d\mathcal{A}}{dt}(\tau)t_W\simeq -\mathcal{A}(\tau+t_W).
\end{eqnarray}
Here we introduced the vector potential $\mathcal{A}(t)$, related to the field force as  
$\mathcal{E}(\tau)=-\frac{d\mathcal{A}}{dt}(\tau)$, i.e.,
\begin{equation}
	A(\tau)=\int_\tau^{\infty}\mathcal{E}(t)dt.
\end{equation}
Thus we arrived at the common formula for attosecond streaking. In correspondence with it, the plot of the momentum of the registered electron versus time mimics the plot of the vector potential with the horizontal shift, equal to the Wigner time delay. 

In a system with Coulomb potential, assuming again that  $\frac{d\mathcal{E}}{dt}(t\to\tau)\to 0$ to avoid the necessity to know the particle trajectory near the center, we can use the approximation $r^{(0)}(t-\tau)\simeq k\{t-\tau-t_{D}[k(t-\tau)]\}$. Integrating by parts and using the fact that $t_D(r)=t_0+t_{CA}(r)$ and $t_{CA}(0)=0$, we get
\begin{eqnarray}
\Delta p&=&
\int_\tau^{\infty}\frac{d\mathcal{E}}{dt}(t)\{t-\tau-t_{D}[k(t-\tau)]\}dt \simeq -\mathcal{A}(\tau+t_0)+\Delta p_{CL}. \label{Dp_C}
\end{eqnarray}
Here the notation is introduced
\begin{eqnarray}
\Delta p_{CL}&=&-\int_\tau^{\infty}\frac{d\mathcal{E}}{dt}(t)t_{CA}[k(t-\tau)]dt=-\mathcal{E}(\tau)t_{CA}(0)+\int_\tau^{\infty}\mathcal{E}(t)\frac{dt_{CA}[k(t-\tau)]}{dt}dt
=\nonumber\\
&=&\int_0^{\infty}\mathcal{E}(t+\tau)\left[\frac{1}{p^{(0)}(kt)}-\frac{1}{k}\right]kdt. \label{Dp_CL}
\end{eqnarray}
The change of the momentum  $\Delta p_{CL}$, arising under the joint action of Coulomb and laser field, is usually referred to as Coulomb–laser coupling \cite{Smirnova2007,Trumm2010,Smirnova2011}
. However, it is easily seen that if one substitutes the step function (\ref{step_Et}) for the field strength in Eq. (\ref{Dp_CL}), then Eq.(\ref{Dp_CL}) with the use of Eq. (\ref{tr_common}) yields $\Delta p_{CL}=\mathcal{E}_0t_{CA}[k(T-\tau)]$, as a result of which Eq.(\ref{Dp_C}) turns into Eq.(\ref{Dp_C_step_Et}). That is, the correction $\Delta p_{CL}$ is connected with the %
Coulomb advance  $t_{CA}(r_{eff})$ of the particle arrival at the point separated by the distance $r_{eff}=k(T-\tau)$ from the center, which, in turn, represents the distance, to each the particle can reache before the strong variation of the laser field. 
For a periodic field with the frequency $\omega$ the characteristic time of the field variation has the order $\sim 1/\omega$, from which it follows that yields an estimate $r_{eff}\sim k/\omega$. 

Let the radiation pulse have the form   $\mathcal{E}(t)=\mathcal{E}_0(t)\sin(\omega t)$, where $\mathcal{E}_0(t)$ is a slowly varying envelope function. Using the expression for the sinus of a sum, and assuming the integrand in Eq.(\ref{Dp_CL}) to decrease much faster, than the variation of $\mathcal{E}_0(t)$, so that the approximation $\mathcal{E}_0(\tau+t)\approx \mathcal{E}_0(\tau)$ is valid, we obtain from Eq.(\ref{Dp_CL}) the following expression:  
\begin{eqnarray}\Delta p_{CL}&=&\frac{\mathcal{E}_0(\tau)}{\omega}\sin(\omega \tau)I_{\cos}(a)+\frac{\mathcal{E}_0(\tau)}{\omega}\cos(\omega \tau)I_{\sin}(a). \label{Dp_CLsincos}
\end{eqnarray}
Here we use the notation 
\begin{eqnarray}
I_{\cos}(a)&=&\int_0^{\infty}\cos(t)\left[\frac{1}{\sqrt{1+2a/t}}-1\right]dt;\\
I_{\sin}(a)&=&\int_0^{\infty}\sin(t)\left[\frac{1}{\sqrt{1+2a/t}}-1\right]dt,
\end{eqnarray}
where the parameter $a$ is defined as  
\begin{equation}
	a=\frac{Z\omega}{k^3}. \label{eq_parameter_a}
\end{equation}

Note that Eq. (\ref{Dp_CLsincos}) may be considered as the result of splitting of the field strength  $\mathcal{E}(t)$ into two terms, $\mathcal{E}(t)=\mathcal{E}_\text{I}(t)+\mathcal{E}_\text{II}(t)$, where the first term $\mathcal{E}_\text{I}(t)=\mathcal{E}(\tau)\sin[\omega(t-\tau)]$ (that yields the first term in Eq.(\ref{Dp_CLsincos})), satisfies the condition $\frac{d\mathcal{E}_\text{I}}{dt}(t=\tau)=0$, which we use to avoid the dependence of the final result on the details of the particle behavior in the vicinity of the center.   

Under the condition $a\ll 1$ the integrals have simple expressions
\begin{eqnarray}
I_{\cos}(a)&\approx& -a\left[\ln\frac{2}{a}-1-\gamma\right]-\frac{3\pi}{4}a^2;\\
I_{\sin}(a)&\approx& -\frac{\pi}{2}a+\frac{3}{2}a^2\left[\ln\frac{2}{a}-\frac{1}{6}-\gamma\right],
\end{eqnarray}
where  $\gamma=0.57721\ldots$ is the Euler constant. If the variation of the envelope function  $\mathcal{E}_0(t)$ during the period $T_{IR}=2\pi/\omega$ is small and  $\mathcal{E}_0(t\to\infty)=0$, then the vector potential is approximately expressed as  $\mathcal{A}(t)\approx \frac{\mathcal{E}_0(t)}{\omega}\cos\omega t$. With 
this fact taken into account, Eq.(\ref{Dp_CLsincos}) can be rewritten as 
\begin{eqnarray}
\Delta p_{CL}&=&-\frac{d\mathcal{A}}{dt}t_{CA}(r_{eff})-\mathcal{A}(\tau)\alpha, \label{Dp_CLfin}
\end{eqnarray}
where 
\begin{eqnarray}
r_{eff}=\frac{k}{\omega}\exp\left[-\gamma+\frac{3\pi}{4}a\right],\label{eq_reff}
\end{eqnarray}
is the effective radius and
\begin{eqnarray}
\alpha=-I_{\sin}(a)=\frac{\pi}{2}a-\frac{3}{2}a^2\left[\ln\frac{2}{a}-\frac{1}{6}-\gamma\right].\label{eq_alpha}
\end{eqnarray}
Finally, the increment of the momentum is expressed as
\begin{eqnarray}
\Delta p&=& -(1+\alpha)\mathcal{A}(\tau+t_S), \label{Dp_C_sin_Et}
\end{eqnarray}
where 
\begin{eqnarray}
t_S=\frac{t_D(r_{eff})}{1+\alpha}\label{eq_tS}
\end{eqnarray}
is the time shift.

We recall that $t_D(r_{eff})$ arises from the first term in Eq.(\ref{Dp_CLsincos}) and the increment of the amplitude of pulse variation $\alpha$ appears from the second one. The equation (\ref{Dp_CLsincos}) follows from the decomposition of the periodic field oscillation into two components, one of them being constant at the ionization time moment  $\tau$ and the other one changing at this moment. Summarising these facts, we can conclude, that, in full analogy with the above model example of a stepwise field, the time delay $t_D(r_{eff})$ is recorded at the moment of the field switch-off, and $r_{eff}$ is the distance reached by the particle at this moment. The field component, changing at the moment of ionization, gives rise to a distortion of this picture, increasing the momentum variation, synchronous with the vector potential of the external field. Against the background of this increased "swing" of the momentum the observed time delay $t_S$ is reduced.  

\begin{figure}[ht]
\includegraphics[angle=-90,width=0.65\textwidth]{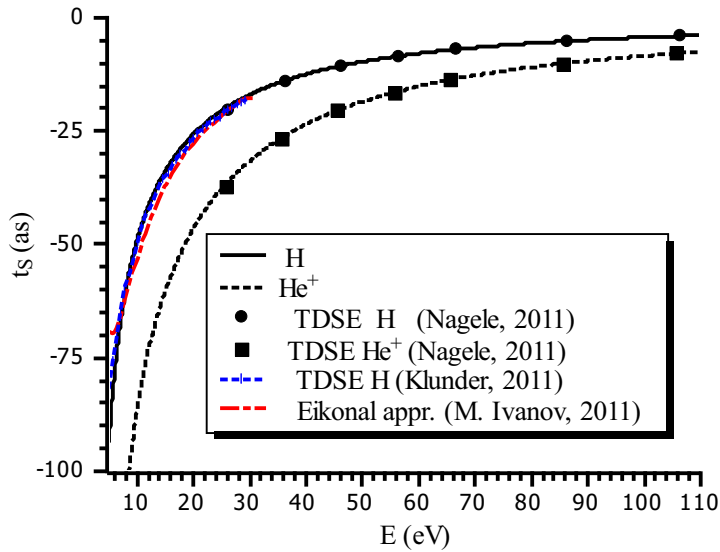}
\caption{(Color online) Dependence of  $t_S$ upon the energy of the ejected electron with the angular momentum  $\ell=1$ for the ionization of a hydrogen atom (solid line) and a helium ion  He$^+$ (dashed line). For the frequency of the IR field $\omega=1.55$ eV. the results are calculated analytically using Eqs.  (\ref{eq_tS}), (\ref{eq_reff}), (\ref{eq_alpha}), and (\ref{eq_parameter_a}) (lines); for comparison the results of solving the time-dependent Schr\"odinger equation for an atom in the field of a femtosecond IR laser pulse from  \cite{Nagele2011} (points) and \cite{Klunder2011} (dotted line) are shown, as well as the result of the eikonal approximation  \cite{Smirnova2011}. \label{FIG_tS}}
\end{figure}
From Fig. \ref{FIG_tS} it is  seen that the simple analytical expression  (\ref{eq_tS}) provides complete agreement with the results \cite{Nagele2011} and \cite{Klunder2011}, obtained from the solution of time-dependent Schr\"odinger equation for an electron affected by the ion field and the field of an IR laser pulse. 

Let us explicitly express the time shift in terms of the Wigner time 
\begin{eqnarray}
t_S=\frac{t_W}{1+\alpha(\omega Z/k^3)}+t_{IR},\label{eq_tS_via_tW}
\end{eqnarray}
where the additive part of the contribution from the Coulomb-laser coupling to the observed time shift is    
\begin{eqnarray}
t_{IR}=-\frac{1}{1+\alpha(\omega Z/k^3)} \frac{Z}{k^3}\left[\ln\frac{2k^2}{\omega}-1-\gamma+\frac{3\pi}{4}\frac{\omega Z}{k^3}\right].\label{eq_tIR}
\end{eqnarray}

\begin{figure}[ht]
\begin{center}
\includegraphics[angle=-90,width=0.5\textwidth]{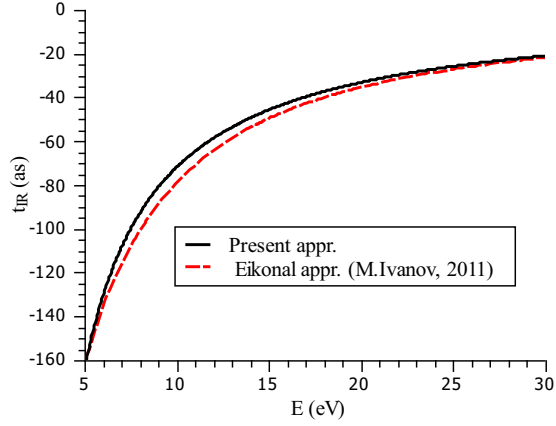}
\end{center}
\caption{(Color online) Dependence of Coulomb-laser coupling contribution $t_{IR}$ to the time shift upon the energy of the electron: our formula (solid line) and the formula, derived from the eikonal approximation \cite{Smirnova2011} (dashed line).\label{FIG_tIR}}
\end{figure}

Using the eikonal approximation, the authors of \cite{Smirnova2011} also derived an analytical expression for $t_{IR}$. Although the dominant term in this expression, logarithmically depending upon  $1/\omega$, is analogous to the dominant term in our Eq. (\ref{eq_tIR}), the minor terms are not similar, and even the power of the momentum that enters them is different.  At $k\to\infty$ the formula from \cite{Smirnova2011} differs from our one by  $2\gamma/k^3$. From Fig.\ref{FIG_tIR} it is seen that our Eq.  (\ref{eq_tIR}) and the formula from \cite{Smirnova2011} yield close but not identical results. For $E=10$ eV the difference between the results amounts to 7 attoseconds, which is available for the present-time experimental facilities. In Fig. \ref{FIG_tS} one can see that our results are apparently closer to those obtained by means of the TDSE numerical solution than the result of  \cite{Smirnova2011}. Yet more important is the difference in the approach and the interpretation of the results. In  \cite{Smirnova2011} $t_{IR}$ was obtained by calculating the initial electron position $r_0$. Naturally, this makes it hard to generalize the results of  \cite{Smirnova2011} over potentials, strongly different from the Coulomb one near the center. We were basing on the interpretation of attosecond streaking as an equivalent
of a virtual detector placed at the large distance $r_{eff}$ from the center, which allows considering only the classical behavior of the electron far from the nucleus. The Coulomb-laser coupling in this interpretation is a record of the Coulomb advance of the particle arrival at the point $r_{eff}$. 

\section{Time delay for spheroidal two-center Coulomb waves}\label{sec:TwoCoulomb}

In this Section and the following ones we consider the time delay of the ionization of two-center targets and the specific phenomena that occur in this case. As a benchmark object we choose the simplest molecule, the molecular hydrogen ion H$_2^{+}$, for which in spheroidal coordinates the separation of variables in the time-independent Schr\"odinger equation is possible 
\[
\xi=\frac{|\mathbf{r}-\frac{\mathbf{R}}{2}|+|\mathbf{r}+\frac{\mathbf{R}}{2}|}{R}
\in [1,\infty );\,\,\,
\eta=\frac{|\mathbf{r}-\frac{\mathbf{R}}{2}|-|\mathbf{r}+\frac{\mathbf{R}}{2}|}{R}
\in [-1,1];\,\,\, \phi \in [0,2\pi).
\]
and the solutions (Coulomb spheroidal functions) are known  \cite{Serov2002,Serov2005}. 

The two-center Coulomb continuum wave function may be presented as a sum of partial spheroidal waves \cite{Serov2002}
\begin{eqnarray}
\varphi_{\mathbf{k}}^{(-)}(\mathbf{r})=(2\pi)^{3/2}4\pi\sum_{lm}\Upsilon_{lm}^*\left(\frac{kR}{2},\theta_e,\phi_e\right)
i^{l}e^{-i\delta_{lm}}T_{ml}\left(\frac{kR}{2},\xi\right)\Upsilon\left(\frac{kR}{2},\arccos\eta,\phi\right).\label{spheroidal_varphi}
\end{eqnarray}
Here the spheroidal harmonics are introduced as 
\begin{equation}
\Upsilon_{lm}(c,\theta,\varphi)=\overline{S}_{ml}\left(c,\cos\theta\right)
\frac{\exp(im\varphi)}{\sqrt{2\pi}};\,\,\,\Upsilon_{lm}(0,\theta,\varphi)=Y_{lm}(\theta,\varphi),\label{Upsdef}
\end{equation}
where $\overline{S}_{ml}\left(c,\eta\right)$ is the normalized quasiangular spheroidal function \cite{Komarov1976}, $m$ is the quantum number of the angular momentum projection onto the molecular axis, $l$ is the quasiazimuthal quantum number, $c=kR/2$ is the unitless parameter.
The quasiradial Coulomb spheroidal function  $T_{ml}(c,\xi)$ has the asymptotic form  
\begin{equation}
T_{ml}(c,\xi\to\infty)=\frac{1}{c\xi }\sin \left( c\xi
+\frac{RZ_+}{2c}\ln(2c\xi)-\frac{l\pi}{2}+\delta_{lm}\right),
\end{equation}
where $Z_+$ is the total charge of the nuclei, $\delta_{lm}$ is the partial wave phase, entering Eq.(\ref{spheroidal_varphi}).
 
First let us consider the case, when after the ionization the electron appears in the state with the fixed angular spheroidal quantum numbers  $m$ and $l$. In analogy with the spherical case (i.e., Eq.(\ref{tauW_ell})) one can introduce the Wigner time delay
for a spheroidal partial wave
\begin{eqnarray}
t_W=\frac{d\delta_{lm}}{dE}. \label{tauW_lm}
\end{eqnarray}
The ejection delay $t_0$ for the given Wigner time delay 
is calculated using Eq.(\ref{eq_t0}), assuming $Z=Z_+=2$, i.e., in this case $t_0$ is in fact the delay compared to the classical electron, ejected from the center of a helium ion He$^+$.

\begin{figure}[ht]
\begin{center}
\includegraphics[angle=-90,width=0.5\textwidth]{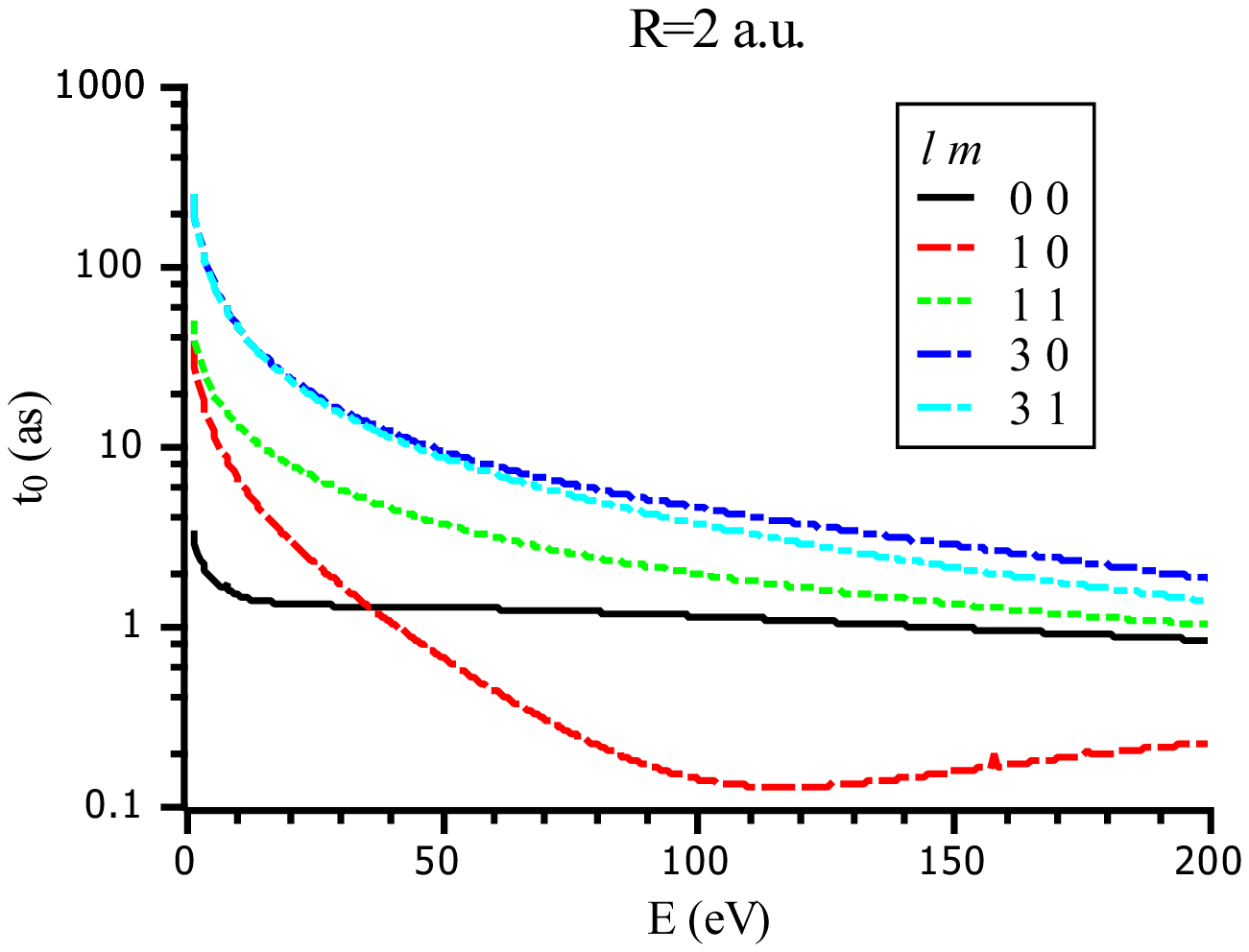}
\\(a)\\
\includegraphics[angle=-90,width=0.5\textwidth]{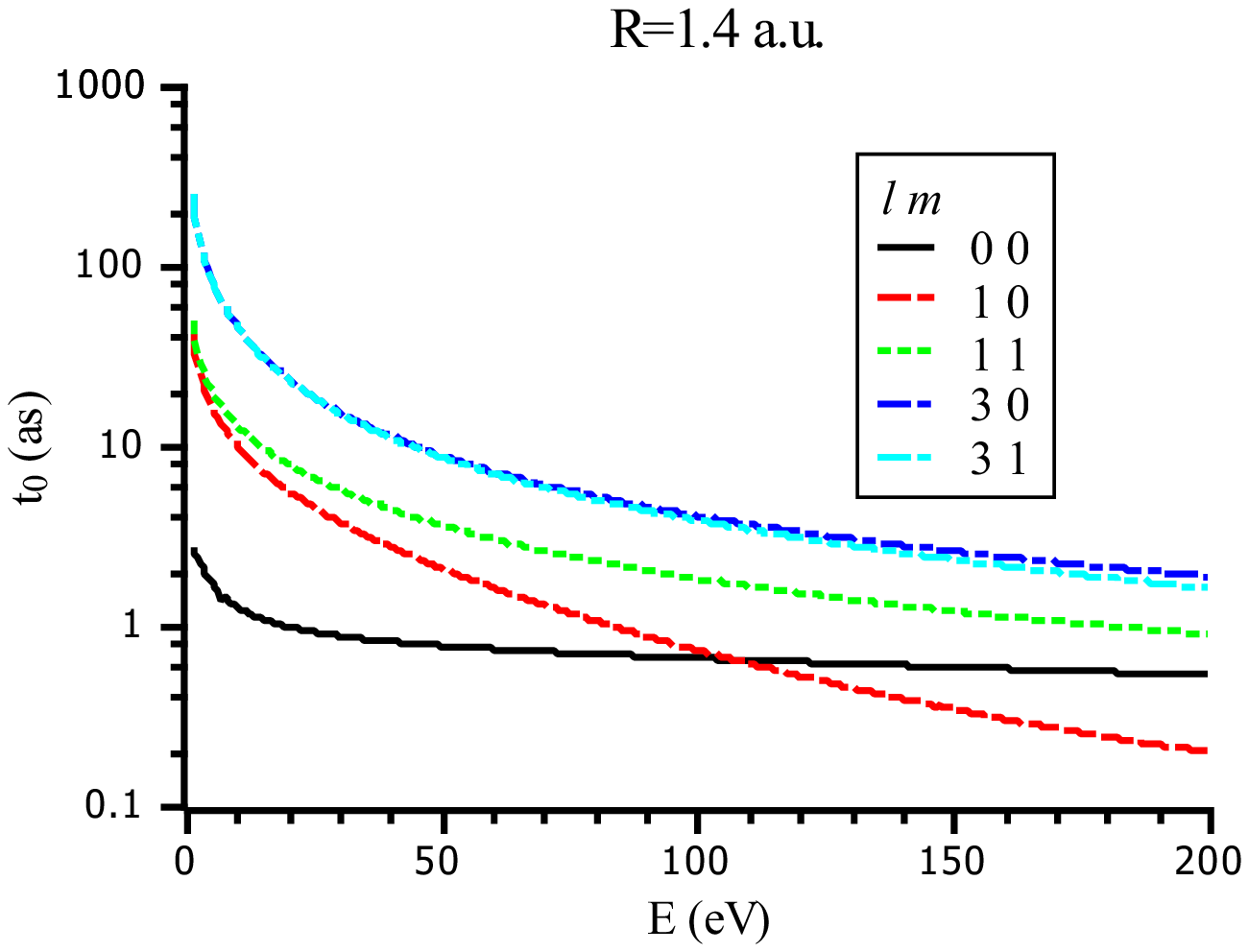}\\(b)
\end{center}
\caption{(Color online) Dependence of the ejection time delay for partial waves of the contnuum  in the hydrogen molecular ion upon the energy of the ejected electron : a) $R=2$ a.u.; b)$R=1.4$ a.u.\label{FIG_H2plus_t0LM}.}
\end{figure}

Fig.\ref{FIG_H2plus_t0LM} presents the dependence of the ejection time delay upon the energy for different partial waves of the H$_2^+$ continuum for two values of the internuclear distance,    $R=2$ a.u. and $R=1.4$ a.u. The distance $R=2$ a.u. is the equilibrium internuclear distance for  H$_2^+$. The distance $R=1.4$ a.u. coincides with the equilibrium internuclear distance in the  hydrogen molecule H$_2$, which makes it easier to compare the results for  H$_2^+$ and H$_2$. 

From Fig.\ref{FIG_H2plus_t0LM} it is obvious that for small energies and any spheroidal quantum numbers $l$ the time $t_0$ tends to that of the helium atom with the same spherical azimuthal quantum numbers. The cause is clear: at very large wavelength the particular form of the potential near the center becomes unimportant, and the main contribution comes from the region, where the potential is practically coincident with the one-center Coulomb one.  At large energies essential differences from the one-center case arise. First, growing difference appears between the curves with the same
$l$, but different $m$. For $m=1$ the deviation from the one-center case is smaller, and  $t_0$ is greater, than for $m=0$. Apparently, this is because the electrons with larger  $m$ move farther from the axis of the molecule. Moreover, the curve for $l=1, m=0$ quickly descends with growing $E$, due to which it crosses the curve for $l=0$, and becomes lower than it. The explanation is that the spheroidal harmonic $\Upsilon_{10}(\eta,\phi)$ possesses a node in the plane, perpendicular to the molecular axis and crossing it in the middle between the nuclei. This node shifts the probability density closer to the nuclei compared with $\Upsilon_{00}(\eta,\phi)$, which has no nodes. Thus, in the case $l=1, m=0$ the electron moves through the regions, where the average attractive potential of the nuclei is larger than in the regions, where the electron with $l=0$ moves, and at large energies this effect dominates over the decelerating effect of the centrifugal potential (which is stronger at  $l=1$). As a result the electron leaves the molecule faster.
 
\begin{figure}[ht]
\begin{center}
\includegraphics[angle=-90,width=0.5\textwidth]{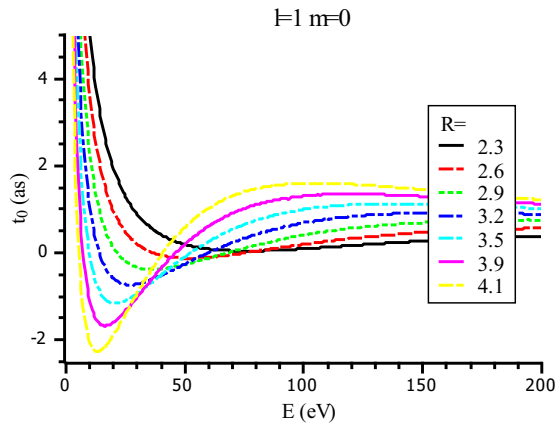}
\end{center}
\caption{(Color online) Dependence of the time delay for the patial wave with $l=1$ and $m=0$ of the H$_2^+$ continuous spectrum upon the energy of the ejected electron at large  $R$\label{FIG_H2plus_t010}.}
\end{figure}

Note, that for $R=2$ a.u. the effect of "advance" of the partial wave with $l=1, m=0$ is more pronounced. Therefore, it seems interesting to trace the variation of the curve for $l=1, m=0$ under the increase of $R$. Figure \ref{FIG_H2plus_t010} shows the dependences  $t_0(E)$ for different values of $R$. It is seen that with the growth of $R$ the dip in $t_0(E)$ is shifted towards smaller $E$ and becomes deeper. At $R>2.6$ a.u. the minimal $t_0(E)$ becomes negative, i.e., the electron in the field of two centers, each having the charge $Z=1$ (only one of them approached close enough) arrives faster, than in the field of one center with the charge $Z=2$. A more consistent interpretation of this effect is the shift of the ejection point. 

Thus, the parameter $t_0$ carries information about the potential in the region, through which the electron has travelled. Since the angular distribution of the differential cross-section at the infinity represents a picture of the angular distribution of these regions, the comparison of $t_0$ for different partial waves can provide information of the target structure. However, in contrast to the one-center case, when the ejection of a particle with the fixed angular momentum $\ell$ is provided by the angular momentum selection rules (i.e., the angular momentum conservation law), for the spheroidal quantum numbers only the parity selection rule holds. Due to this fact the experimental measurement of time delays for partial waves is a nontrivial problem.

\section{Time delay of H$_2^+$ ionization}\label{sec:H2plus}
Since in the process of single-photon ionization of a two-center target, e.g.,  H$_2^+$, the dipole transition generally brings the system to a state, containing more than one partial spheroidal wave, the time delay is expected to depend upon the angle. To avoid the problem with possible jumps of the calculated phase by $2\pi$ (due to the phase uncertainty of a periodic function) and by $\pi$ (near the resonances), instead of preliminary calculation of the phase followed by its differentiation, it is convenient to calculate directly the phase derivative as the imaginary part of the logarithmic derivative of the ionization amplitude
\begin{eqnarray}
t_W=\Im\left(\frac{1}{f}\frac{\partial f}{\partial E}\right).\label{dlog_f}
\end{eqnarray}

\begin{figure}[ht]
\begin{center}
\includegraphics[angle=-90,width=0.5\textwidth]{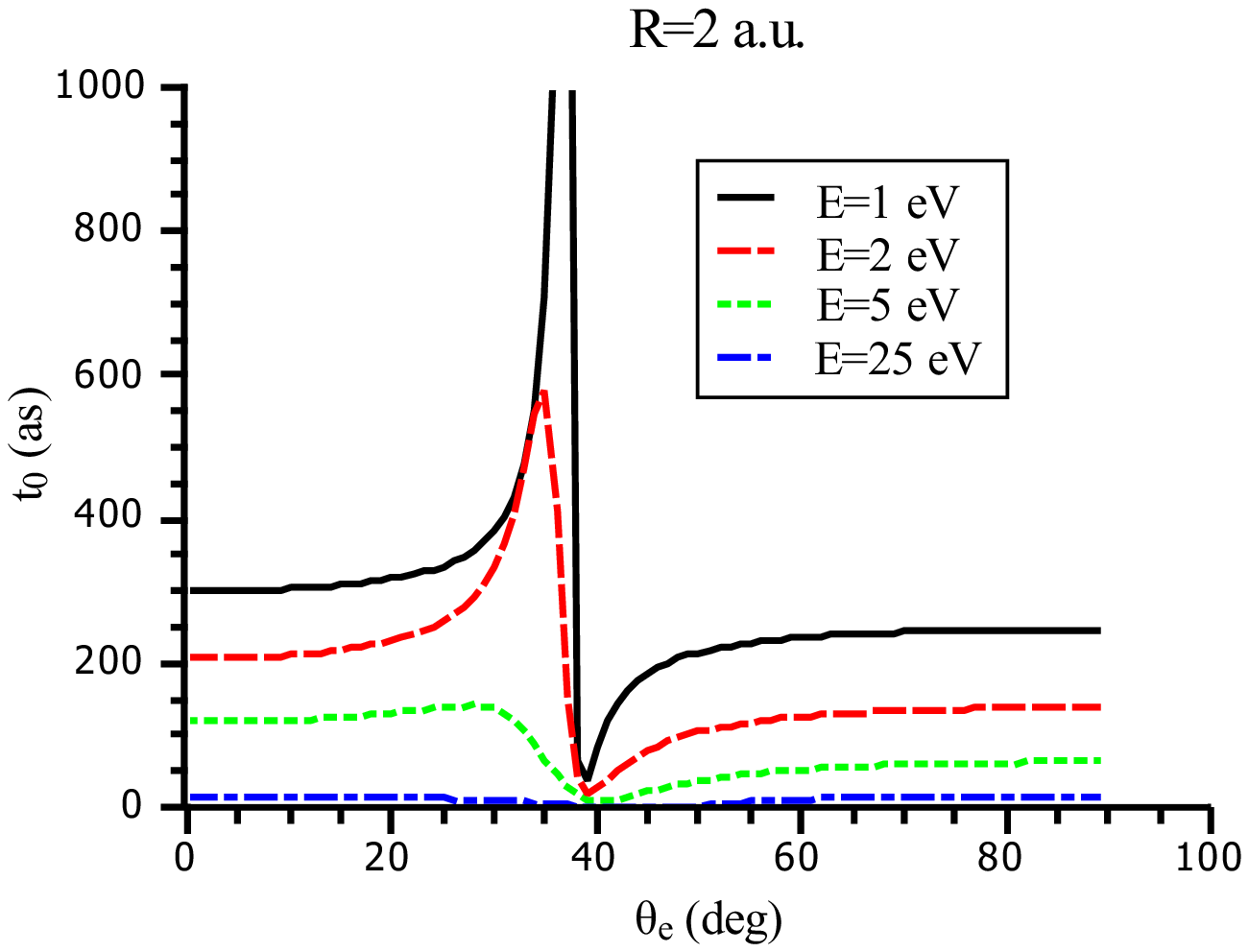}
\\(a)\\
\includegraphics[angle=-90,width=0.5\textwidth]{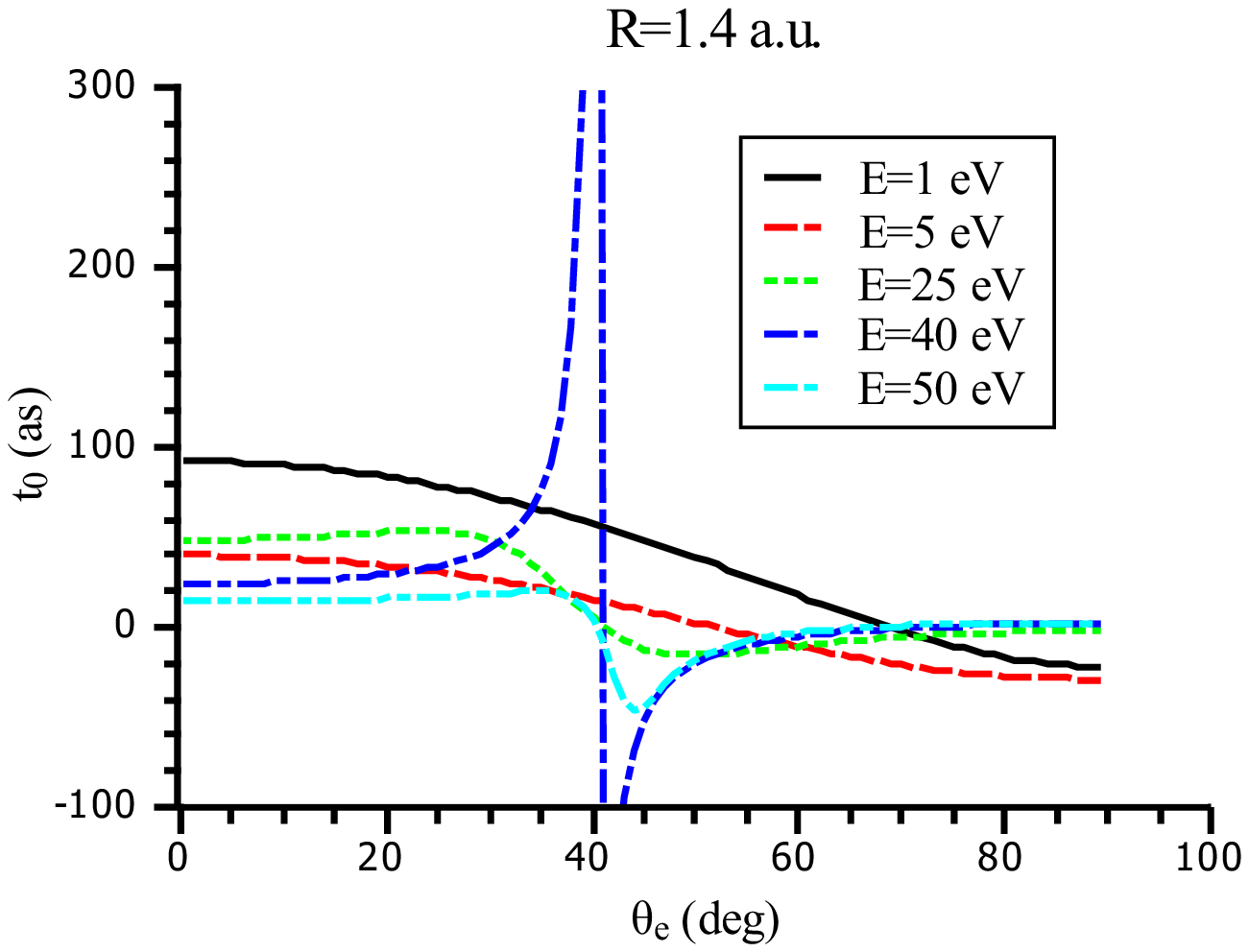}\\(b)
\end{center}
\caption{(Color online) Dependence of the time delay upon the ejection angle for the ionization of molecular hydrogen ion in the ground state at $\mathbf{e}\parallel \mathbf{R}$: a) $R=2$ a.u.; b)$R=1.4$ a.u.\label{FIG_H2plus_t0theta}.}
\end{figure}

Figure \ref{FIG_H2plus_t0theta} presents the angular distributions of the ejection delay $t_0$ when H$_2^+$ is ionized in the initial ground state with the molecular axis parallel to the polarization of the ionizing field, for different values of energy and internuclear distance. It is seen that, similar to H$_2$ (Fig.\ref{FIG_H2_Wigner_delay}), the behavior of $t_0(\theta_e)$ demonstrates characteristic "zigzags". The plots of the differential cross-section demonstrate dips at the same positions  
\cite{IvanovSerov2012}. At some energy ($E\sim 0$ eV for $R=2$ a.u., and $E\approx 40$ eV for $R=1.4$ a.u.) the swing of these "zigzags" of $t_0(\theta_e)$ becomes huge. 

\begin{figure}[ht]
\begin{center}
\includegraphics[angle=-90,width=0.5\textwidth]{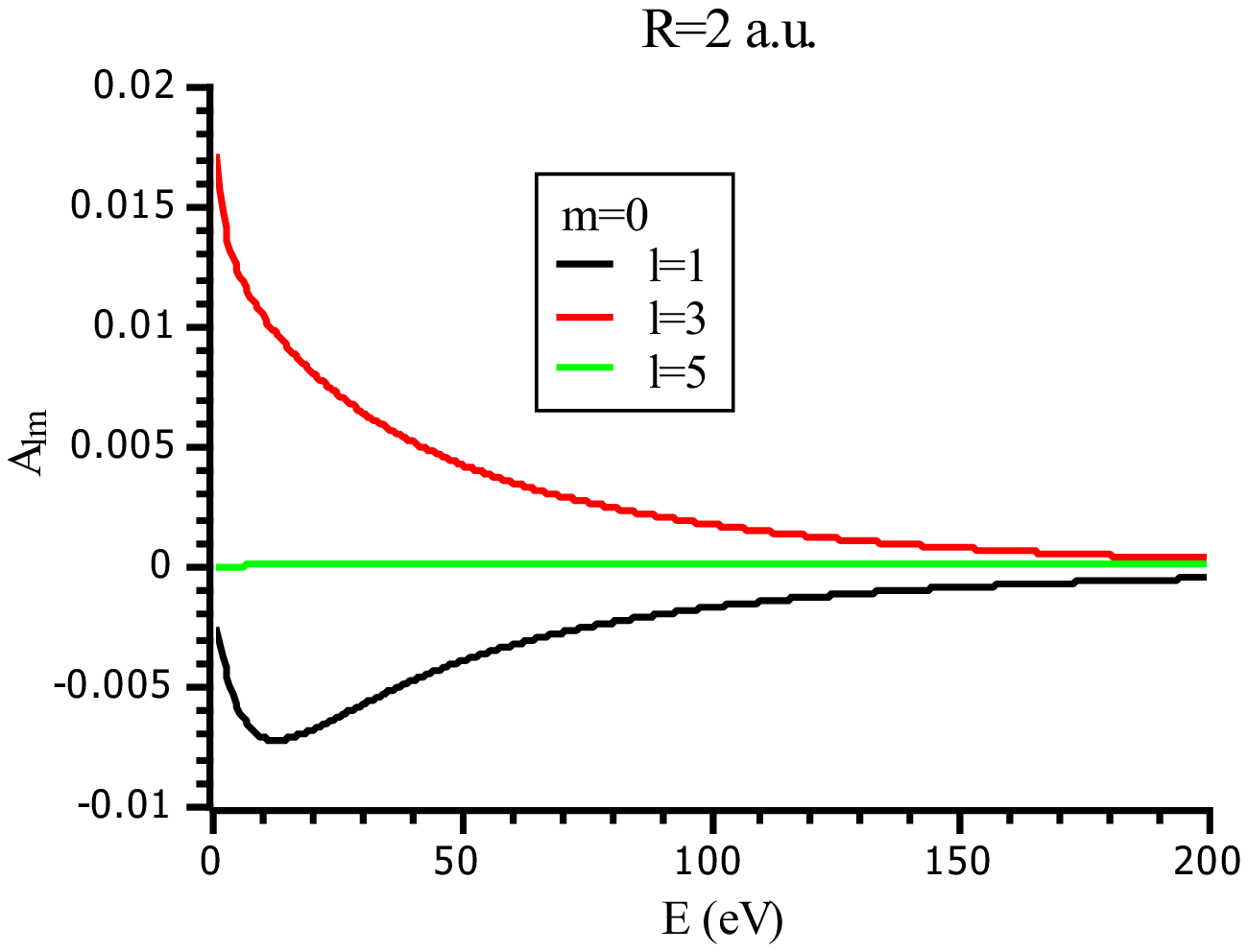}
\\(a)\\
\includegraphics[angle=-90,width=0.5\textwidth]{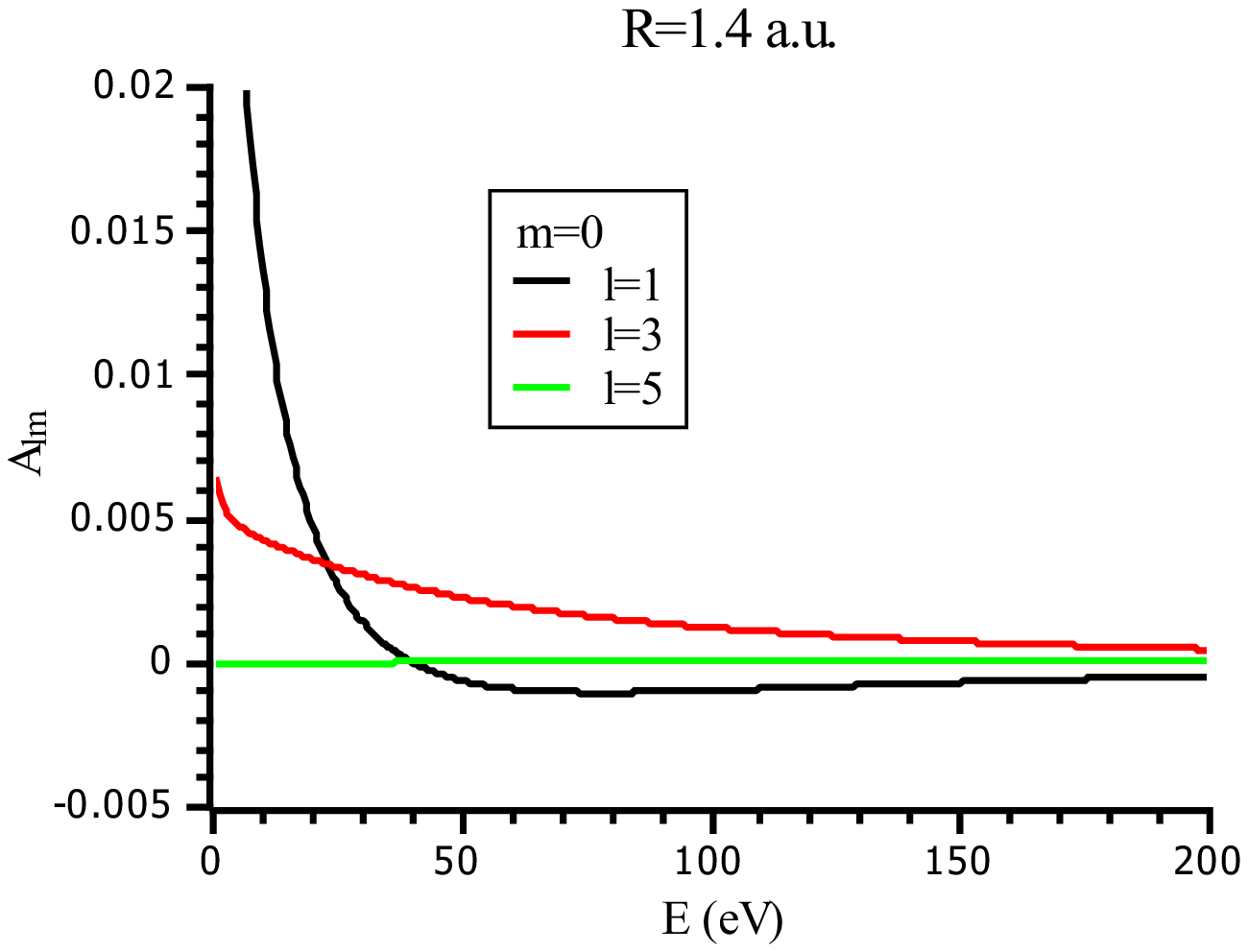}\\(b)
\end{center}
\caption{(Color online) Dependence of the partial components of the ionization amplitude  for the hydrogen molecular ion in the ground state at $\mathbf{e}\parallel \mathbf{R}$: a) $R=2$ a.u.; b)$R=1.4$ a.u.\label{FIG_H2plus_ALM}.}
\end{figure}

Since the delays for partial waves (Fig. \ref{FIG_H2plus_t0LM}) have no singularities, it is apparent that the latter arise from superposition of several partial waves. Eq. (\ref{spheroidal_varphi}) allows the following expression of the ionization amplitude having angular dependence 
\begin{eqnarray}
f(\mathbf{k}_e)=\sum_{lm}A_{lm}i^{-l} e^{i\delta_{lm}}\Upsilon_{lm}(c,\theta_e,\phi_e),\label{f_expan_spheroid}
\end{eqnarray}
where $A_{lm}$ is the amplitude of the transition to the spheroidal partial wave with the quantum numbers  $l,m$. Figure \ref{FIG_H2plus_ALM} shows the amplitudes  $A_{l0}$ of the dipole transition from the ground state of  H$_2^+$ to the partial waves of the continuum depending of the final energy. It is seen that for all energies only the partial waves with  $l=1$ and $l=3$ are essential. 

Consider the ionization amplitude as a sum of contributions from two dominating partial waves
\begin{eqnarray}
f=A_{10}e^{i\delta_1-i\pi/2}\Upsilon_{10}+A_{30}e^{i\delta_3-i3\pi/2}\Upsilon_{30}. \label{f_A1A3}
\end{eqnarray}
In correspondence with Eq.(\ref{dlog_f}), $t_W$ tends to infinity near the nodes $f(\theta_{e0},\phi_{e0})=0$, if for the same angles  $\frac{\partial}{\partial E}f(\theta_{e0},\phi_{e0})\neq 0$. Obviously, the sum of two functions with different position of nodes can be zero only in two cases: (i) the phases of partial waves satisfy the condition  $\delta_{30}-\delta_{10}=N\pi$, where  $N$ is an integer; (ii) one of the expansion coefficients for the given energy turns into zero, so that the sum actually consists of a single spheroidal harmonic.
Figure  \ref{FIG_H2plus_ALM} shows that the second of these statements is valid, i.e., there are such values of energy ($E\sim 0$ eV at $R=2$ a.u. and $E=39$ eV at $R=1.4$ a.u.), at which the amplitude for $l=1$ turns into zero. Just at these energies Fig.  \ref{FIG_H2plus_t0theta} demonstrates the singularity of $t_0(\theta_e)$. 

The situation, when a partial amplitude turns into zero at a certain value of energy, is commonly referred as Cooper's minimum, in honor of J.W. Cooper, who theoretically predicted this phenomenon in the photoionization of noble gases  \cite{Cooper1962}). It was shown \cite{Semenov2003} that in the case of H$_2$ photoionization the square modulus of the spherical  $p$-wave has a minimum at $\omega\approx 82$ eV (energy $E\approx 66$ eV), so that in the vicinity of this energy the contribution of  $f$-wave is dominant. 

In the case of H$_2^+$ ionization, considered here, the contribution that turns into zero comes from the spheroidal wave with the quantum number $l=1$, rather than from the spherical $p$-wave. The remaining wave is the spheroidal wave  with  $l=3$, so it is more consistent to refer the phenomenon as ``spheroidal Cooper's minimum''.

A spheroidal harmonic can be expanded over the spherical ones  \cite{Komarov1976} 
\begin{eqnarray}
\Upsilon_{lm}(c,\theta,\phi)=\sum_{\ell}d^{lm}_\ell(c)Y_{\ell m}(\theta,\phi),
\end{eqnarray}
and correspondingly, the expansion over spheroidal harmonics \ref{f_expan_spheroid}) can be transformed into the expansion over the spherical ones. If  $c=kR/2$ is not very large, the spheroidal harmonic  $\Upsilon_{30}$ is rather close to the spherical one,  $Y_{30}$, however,  $\Upsilon_{30}$ contains also the contribution from  $Y_{10}$, so that in the ``spheroidal Cooper's minimum'' the amplitude of the spherical  $p$-wave does not turn into zero.  Moreover, since in the vicinity of the Cooper's minimum the phase difference of spheroidal waves is  $\delta_{30}-\delta_{10}\sim \pi/2$, the amplitude of $p$-does not turn into zero exactly at any energy, although its square modulus has a minimum near the energy of the ``spheroidal Cooper's minimum''. This is the difference between Cooper's minima in two-center systems and in one-center ones, like noble gas atoms \cite{Cooper1962}. However, since the Cooper's minimum for  H$_2^+$ with the equilibrium internuclear distance  $R=2$ a.u. is present at very small energy of the ejected electron, and  $\Upsilon_{30}(c\to 0,\theta,\phi)=Y_{30}(\theta,\phi)$, the photoionization of  H$_2^+$ at small energies is to yield purely ``octupole'' electrons, i.e., those with  $\ell=3$.

Taking the expansion (\ref{f_A1A3}) for energies $E$ in the vicinity of the energy of Cooper's minimum  $E_C$, for which by definition $A_{10}(E_C)=0$, and substituting  it into Eq. (\ref{dlog_f}), it is easy to obtain
\begin{eqnarray}
t_W=\frac{d\delta_{30}}{dE}+\frac{\Upsilon_{10}}{\Upsilon_{30}}\frac{A_{10}'(E_C)}{A_{30}(E_C)}\sin(\delta_{30}-\delta_{10}), \label{Cooper_time}
\end{eqnarray}
where $A_{10}'(E_C)=\left.\frac{dA_{10}}{dE}\right|_{E=E_C}$. From this formula it is evident that the instability of the dependence of Wigner time delay upon the ejection angle at the energy  $E_C$ is by no means related to the phase derivatives of the partial waves, but depends on the the difference of the  phases themselves $\delta_{30}-\delta_{10}$. As to the singularities, they arise at the ejection angles  $\theta_e$, for which the function $\Upsilon_{30}(c,\theta_e,\phi)$ possesses nodes that do not coincide with the node  $\theta_e=90^\circ$ of the function $\Upsilon_{10}(c,\theta_e,\phi)$. 

\begin{figure}[ht]
\begin{center}
\includegraphics[angle=-90,width=0.5\textwidth]{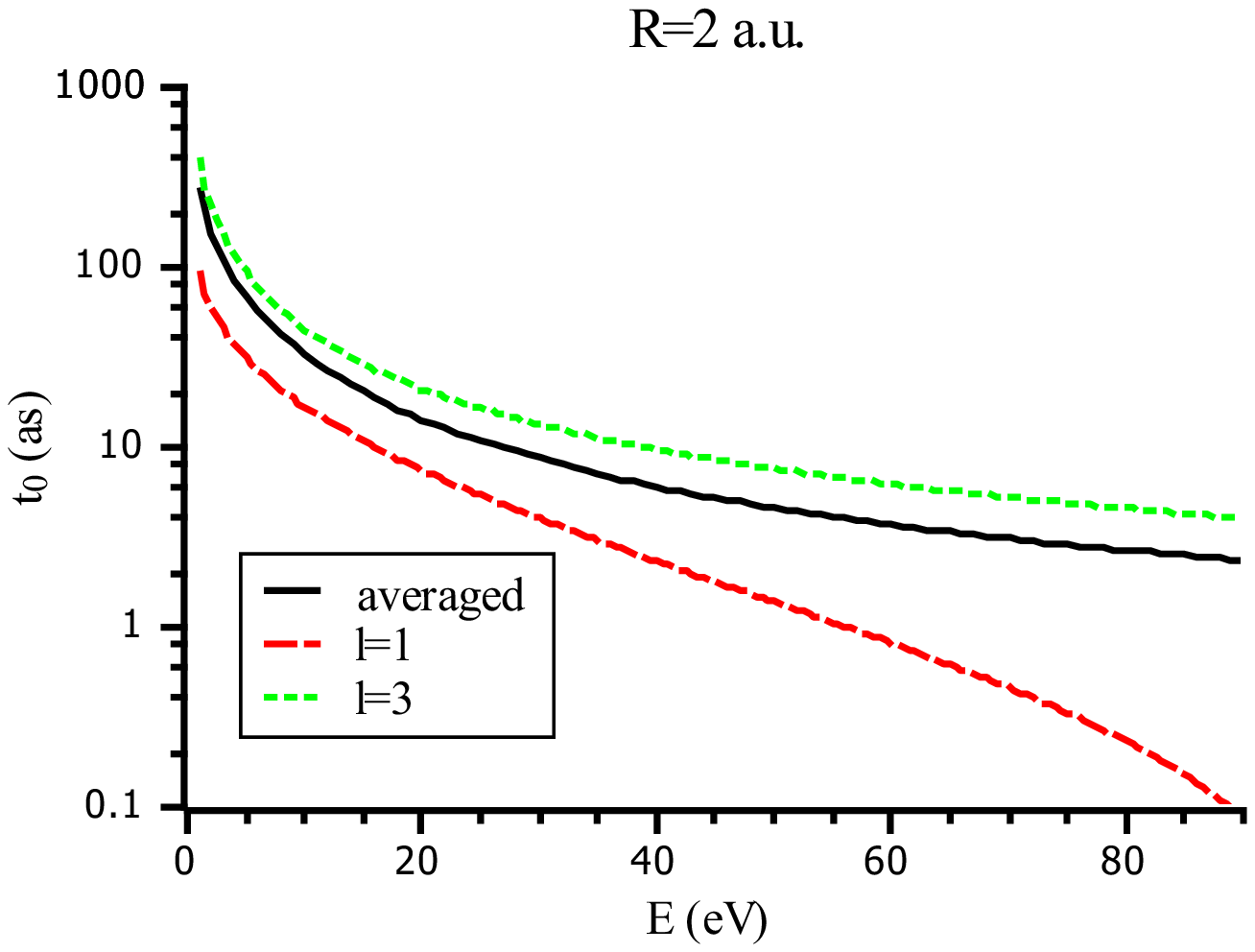}
\\(a)\\
\includegraphics[angle=-90,width=0.5\textwidth]{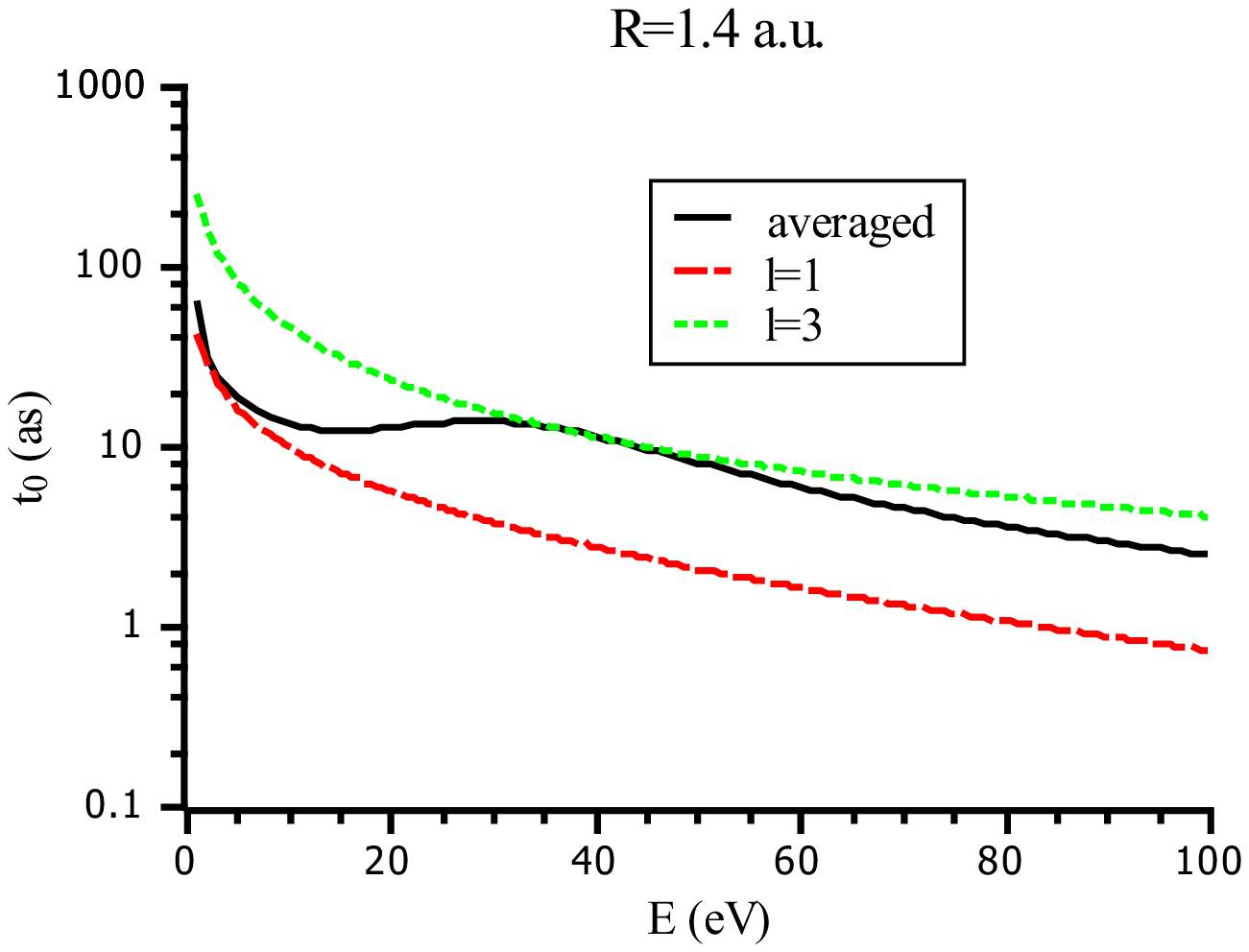}\\(b)
\end{center}
\caption{(Color online) Dependence of the delay $\overline{t}_0$, averaged over the angles, upon the energy of the electron for the ionization of the hydrogen molecular ion in the ground state at   $\mathbf{e}\parallel \mathbf{R}$ (black), as well as that for the delay of partial waves with  $l=1$ (red) and $l=3$ (green): a) $R=2$ ; b)$R=1.4$ a.u.\label{FIG_H2plus_t0aver}.}
\end{figure}

In order to study the angle-independent  part of $t_W$, consider $t_W$, averaged over the directions of ejection
\begin{eqnarray}
\overline{t}_W=
\frac{\oint t_W \sigma^{(1)}d\Omega}{\oint\sigma^{(1)}d\Omega}
=\left[\sum_{lm}|f_{lm}|^2\right]^{-1}\sum_{lm}\Im\left(f_{lm}^*\frac{df_{lm}}{dE}\right).
\end{eqnarray}
where $\sigma^{(1)}=\frac{d\sigma}{d\Omega_e}$ is the ionization differential cross-section, and the partial amplitude $f_{lm}=A_{lm}i^{-l}e^{i\delta_{lm}}$. From this formula it is easy to see that in the Cooper's minimum it should be  $\overline{t}_W=\frac{d\delta_{30}}{dE}$. Indeed, from Fig. \ref{FIG_H2plus_t0aver} and particularly from Fig.  \ref{FIG_H2plus_t0aver}b it is evident that  $\overline{t}_0(E)$ near $E_C$ has a maximum and coincides with the delay of the partial wave  with  $l=3$. 

Note, however, that although $\overline{t}_0(E)$ is convenient for theoretical analysis, its experimental determination presents a difficult problem, since it requires multiple measurements with different orientations of the polarization vector of the ``streaking'' IR radiation with respect to the polarization of ionizing radiation.

\section{Interpretation of singularities in the angular dependence of time delay} \label{sec:Gaussians}

In the previous Section we determined the conditions, under which the singularities in the angular distribution of the time delay occur.  However, their physical meaning remained unclear. Although the huge values of the time delay appear in the vicinity of zeros of the differential cross section, in principle, the time delay can be measured in this case, too. Large negative delays are in apparent contradiction with the causality principle, since from the formal classical viewpoint it means that the electron appears to be ejected before the external action on the molecule. 

It is common to assume that when the molecule is subject to a laser pulse with Gaussian envelop function, and the pulse is long enough, then the momentum distribution of the electrons, ejected as a result of ionization, is also Gaussian  \cite{IvanovSerov2012}. The Gaussian momentum distributon gives results in a Gaussian wave packet at large distances from the center. Following the Ehrenfest theorem, the behavior of the center of a wave packet (even if it is non-Gaussian) coincides with that of the classical particle, so that the classical interpretation of the time delay, used, e.g., in Section \ref{sec:streakingCoulomb}, is correct. But the packet center position itself depends on the packet shape distortions, arising due to quantum effects. Therefore it appears necessary to consider the problem in terms of wave packet motion.

Let the system be a subject to a laser (XUV) pulse having the Gaussian envelope function, the carrier frequency $\omega_0$, and the duration $T$, i.e., the field strength expressed as
\begin{eqnarray}
\mathcal{E}(t)=\mathcal{E}_0 \exp\left[-\frac{t^2}{2T^2}\right]\sin\omega_0t.
\end{eqnarray} 
In the case of weak ionizing field the first-order perturbation theory yields the following expression for the wave function, describing the wave packet of ejected electrons
\begin{eqnarray}
\psi(\mathbf{r},t)\sim \int f(k\mathbf{n}) \exp\left[-\frac{w_0^2(k-k_0)^2}{2}\right]e^{ikr-iEt}dk.
\end{eqnarray} 
Here the initial spatial width of the packet  $w_0=k_0T$, the central momentum $k_0=\sqrt{2(\omega_0-I_1)}$ ($I_1$ being the ionization potential), and $f(\mathbf{k})=\langle \mathbf{k} |z| 1\sigma_g \rangle$ is the amplitude  (\ref{ampl1B}) of the system ionization by a  continuous wave having the frequency $\omega=k^2/2+I_1$ in the first Born approximation.

Consider the dipole matrix element (\ref{f_A1A3}) in the energy range in the vicinity of a Cooper's minimum. Since from Eq. (\ref{Cooper_time}) it is obvious that the partial time delay
determines only the general angle-independent time shift, while we are interested in the interpretation of the second term  (\ref{Cooper_time}), we will assume for simplicity that the derivative of the phase is zero. Near the Cooper's minimum  $A_{10}(E)\simeq A_{10}'(E_C)(E-E_C) \approx A_{10}'(E_C)(k-k_C)k_C$, therefore, from Eq. (\ref{f_A1A3}) it follows that
\begin{eqnarray}
f(\mathbf{k}) \approx -(k-k_C)k_C A_{10}'(E_C) ie^{i\delta_{10}}\Upsilon_{10}(\theta_e,\phi_e)+A_{30}(E_C) ie^{i\delta_{30}}\Upsilon_{30}(\theta_e,\phi_e). 
\end{eqnarray}
Apparently, at the angles, corresponding to zeros of the function  $\Upsilon_{30}(\theta_e,\phi_e)$, the electron wave packet becomes
essentially non-Gaussian, and, therefore, the case of Cooper's minimum requires special consideration.

Consider the electron wave function after ionization in the  $k$-representation 
\begin{eqnarray}
\psi(\mathbf{k})=f(\mathbf{k}) \sqrt{\frac{w_0}{\sqrt{\pi}}}\exp\left[-\frac{w_0^2(k-k_0)^2}{2}\right].
\end{eqnarray}
Introduce the coefficients of its expansion
\begin{eqnarray}
C_n=\int \varphi_n(k-k_0)\psi(\mathbf{k})dk 
\end{eqnarray}
over the basis of functions
\begin{eqnarray}
\varphi_n(k)=\sqrt{\frac{w_0}{2^n n!\sqrt{\pi}}}H_n(ak)\exp\left[-\frac{w_0^2k^2}{2}\right],\label{kBasis}
\end{eqnarray}
where $H_n(x)$ is the Hermite polynomial. The functions (\ref{kBasis}) in the coordinate representation have the form of expanding wave packets
\begin{eqnarray}
\psi_n(x,t)=\sqrt{\frac{1}{2^n n!\sqrt{\pi}w}}H_n(x/w)\exp\left[-\frac{x^2}{2w^2}(1-it/w_0^2)-i\beta_n(t)\right],\label{expandBasis}
\end{eqnarray}
where $w(t)=w_0\sqrt{1+(t/w_0^2)^2}$ is the packet width, growing in time, and the phase is expressed as  $\beta_n(t)=(n+1/2)\arctan(t/w_0^2)-n\pi/2$. The functions $\varphi_n(k-k_0)$ in the coordinate representation equal to $\psi_n(x,t)\exp(ik_0r-i\frac{k_0^2}{2}t)$. Here and everywhere below $x=r-k_0t$. 

If the width $1/w_0$ of the packet momentum distribution is small and $f(\mathbf{k})$ has no singularities and zeros of the order higher than the first, then only two first coefficients of the expansion, $C_0$ and $C_1$, will not be small. It is not difficult to show that in this case the shift of the center of mass of the packet $\langle x \rangle=\langle r \rangle - k_0t$ with respect to the uniform motion with the velocity $k_0$ will amount to
\begin{eqnarray}
\langle x \rangle = 2\frac{\Re[C_1^*C_0 \langle \psi_1|x|\psi_0 \rangle]}{|C_0|^2+|C_1|^2}
\end{eqnarray}
Here the dipole matrix element is expressed as 
\begin{eqnarray}
\langle \psi_1|x|\psi_0 \rangle = \frac{w}{\sqrt{2}}e^{i(\beta_1-\beta_0)}.
\end{eqnarray}
Using the asymptotic expression  $e^{i(\beta_1-\beta_0)}= 1-iw_0^2/t+O(t^{-2})$, we get
\begin{eqnarray}
\langle x \rangle \simeq \sqrt{2}\frac{\Re[C_1^*C_0]}{|C_0|^2+|C_1|^2}\frac{t}{w_0}+\sqrt{2}\frac{\Im[C_1^*C_0]}{|C_0|^2+|C_1|^2}w_0.\label{WPx_aver}
\end{eqnarray}
The first term in this expression grows linearly in time, i.e., actually describes the velocity correction
\begin{eqnarray}
\Delta k=\frac{d\langle x \rangle}{dt}= \frac{\sqrt{2}}{w_0}\frac{\Re[C_1^*C_0]}{|C_0|^2+|C_1|^2}.
\label{Cooper_velocity_gaussian}
\end{eqnarray}
This correction arises simply because the momentum distribution is different from  Gaussian centered at $k=k_0$. Indeed, $\Delta k$ turns into zero if the coefficient describing the deviation from the Gaussian shape $C_1=0$.

The second term in Eq.(\ref{WPx_aver}) describes the delay of the packet center arrival at the point $r$ compared with the arrival time $r/(k_0+\Delta k)$ in the case of uniform rectilinear motion
\begin{eqnarray}
t_{WP}=\frac{\langle x \rangle-\Delta k t}{k_0}= \frac{\sqrt{2}w_0}{k_0}\frac{\Im[C_1^*C_0]}{|C_0|^2+|C_1|^2}. \label{Cooper_time_gaussian}
\end{eqnarray}
Note, that the delay  $t_{WP}$ of the packet center of mass arises due the temporal variation of the packet shape, i.e., actually, because of the fact that the common phases of the wave packet without a node, $\beta_0(t)$, and the packet with a node, $\beta_1(t)$, have different time dependence.  The functions (\ref{expandBasis}) are solutions of the time-dependent Schr\"odinger equation with the Hamiltonian of a harmonic oscillator with variable frequency $\varpi(t)=1/\sqrt{w(t)}$, obtained from the equation for a free particle by transformation to the time-scaled coordinate system \cite{Serov2007} $x/w(t)$.
The phases $\beta_n(t)=-n\pi/2+\int_0^t\epsilon_n(t)dt$ are integrals of the eigenenergies $\epsilon_n=(n+1/2)\varpi$ of  harmonic oscillator with variable frequency $1/\sqrt{w(t)}$ over time. Therefore, we can conclude that the delay $t_{WP}$ is a consequence of the specific quantization that arises when the free particle motion is considered in the time-scaled coordinate system.

Consider the case, when the central frequency of the ionizing laser pulse coincides with a Cooper's minimum, i.e., when  $k_0=k_C$. Then
\begin{eqnarray}
C_0&=&-A_{30}(E_C) i e^{i\delta_{30}}\Upsilon_{30};\\
C_1&=&\frac{1}{\sqrt{2}w_0}k_C A_{10}'(E_C) i e^{i\delta_{10}}\Upsilon_{10}. 
\end{eqnarray}
The corresponding combinations of coefficients that enter the expression  (\ref{Cooper_velocity_gaussian}) and (\ref{Cooper_time_gaussian}) are
\begin{eqnarray*}
|C_0|^2+|C_1|^2&=& |A_{30}(E_C)\Upsilon_{30}|^2+\frac{k_C^2}{2w_0^2}\left|A_{10}'(E_C)\Upsilon_{10}\right|^2;\\
C_1^*C_0&=&\frac{k_C}{\sqrt{2}w_0}A_{30}(E_C)A_{10}'(E_C)e^{i(\delta_{30}-\delta_{10})}\Upsilon_{10}\Upsilon_{30}. 
\end{eqnarray*}

Eq. (\ref{Cooper_time_gaussian}) at $w_0\to\infty$ yields the expression that does not depend upon  $w_0$ and completely coincides with the second term in Eq.(\ref{Cooper_time}). The shift of the central momentum at $w_0\to\infty$ 
\begin{eqnarray}
\Delta k=
\frac{1}{w_0^2}\frac{\Upsilon_{10}}{\Upsilon_{30}}\frac{A_{10}'(E_C)}{A_{30}(E_C)}\cos(\delta_{30}-\delta_{10}),
\end{eqnarray}
i.e., for $w_0=\infty$ the velocity shift turns into zero.
Near the node of the angular distribution, i.e., at  $\Upsilon_{30}\to 0$, the wave packet center time delay
\begin{eqnarray*}
t_{WP}=\frac{2w_0^2}{k_C^2} \frac{A_{30}(E_C)}{A_{10}'(E_C)}\sin(\delta_{30}-\delta_{10})\frac{\Upsilon_{30}}{\Upsilon_{10}}.
\end{eqnarray*}
Thus, the time delay in the vicinity of the node of the cross-section does not tend to infinity in reality. On the contrary, for any final $w_0$ its value tends to zero, or, more rigorously, to the delay of the partial spheroidal wave with $l=1$, neglected in the above consideration for simplicity.

The maxima of $|t_{WP}|$ are attained at $|C_0|=|C_1|$. 
The maximal possible delay or advance is 
\begin{equation}
\max_{\Omega_e}|t_{WP}|= \frac{w_0}{k_0} \frac{\sin(\delta_{30}-\delta_{10})}{\sqrt{2}}=T\frac{\sin(\delta_{30}-\delta_{10})}{\sqrt{2}},\label{tWPmax}
\end{equation}
i.e., it apparently does not exceed the uncertainty of the arrival time of the ionizing laser pulse, equal to its duration $T$. Thus, there is no contradiction with the causality principle.

\section{Conclusion}
It is demonstrated that the known method of attosecond streaking is equivalent to a device, placed at a certain distance from the center and recording the delay of the electron arrival. We show that the laser-Coulomb coupling, arising in the theory of attosecond streaking, is related to the Coulomb advance of the particle arrival at this virtual detector. The energy dependence of the time delay is studied for partial two-center spheroidal Coulomb waves. It is shown that for the energies, coincident with those of the Cooper's minimum (in which the amplitude of one of the partial waves of the ejected electron turns into zero), at angles, coinciding with the nodes of the angular spheroidal function having the quantum number $l=3$, the angular distribution of the Wigner time delay for two-center targets has singularities, in the vicinity of which the time delay takes large positive and negative values. By analyzing the dynamics of the wave packet of the ejected electron it is shown that these singularities do not lead to violation of the causality principle, since in the process of ionizing the molecule by an external laser pulse with finite duration the real delay or advance does not exceed the duration of the laser pulse.

The authors acknowledge support of the work from the Russian Foundation for Basic Research (Grant No. 11-01-00523-a).

\end{document}